\documentclass[%
 reprint,
 amsmath,amssymb,
 aps,
]{revtex4-2}
\usepackage{lipsum}
\usepackage{graphicx}
\usepackage{dcolumn}
\usepackage{bm}
\usepackage{color}
\usepackage{comment}
\usepackage{physics}

\begin{document}
\preprint{APS/123-QED}
\title{{\color{blue} Why the first magic-angle is different from others in  twisted  graphene bilayers: interlayer currents, kinetic and confinement energy and wavefunction localization }}
\author{Leonardo A. Navarro-Labastida, Abdiel Espinosa-Champo,  Enrique Aguilar-Mendez, Gerardo G. Naumis}
\date{November 2021}
\email{naumis@fisica.unam.mx}
\affiliation{%
Depto. de Sistemas Complejos, Instituto de F\'isica, \\ Universidad Nacional Aut\'onoma de M\'exico (UNAM)\\
Apdo. Postal 20-364, 01000, CDMX, M\'exico.
}%
\begin{abstract}
The chiral Hamiltonian for twisted graphene bilayers is analyzed in terms of its squared Hamiltonian which removes the particle-hole symmetry and thus one bipartite lattice, allowing to write the Hamiltonian in terms of a $2\times 2$ matrix. This brings to the front the three main physical actors of twisted systems: kinetic energy, confinement potential and an interlayer interaction operator which is divided in two parts: a  non-Abelian interlayer operator and an operator which contains an interaction energy between layers. Here, each of these components is analyzed as a function of the angle of rotation, as well as in terms of the wave-function localization properties.   In particular, it is proved that the non-Abelian operator represents interlayer currents between each layer triangular  sublattices, i.e., a second-neighbor interlayer current between bipartite sublattices. 
{\color{blue} A  crossover is seen between such contributions and thus the first magic angle is different from other higher order magic angles. Such angles are determined by a balance between the negative energy contribution from interlayer currents and the positive contributions from the kinetic and confinement energies. A perturbative analysis performed around the first magic angle allows to explore analytically the details of such energy balance.}

\end{abstract}

\maketitle
\section{Introduction}

Twisted bilayer graphene (TBG) exhibits unconventional superconducting phases and Mott insulating states \cite{Cao2018}. Such discovery was made by working upon previous theoretical efforts which suggested a path to increase many body interactions \cite{MacDonald2011,Santos2007,Guinea2012}.  In particular, MacDonald et. al. \cite{MacDonald2011} found that at certain twisting angles, TBG presents flat-bands where the Fermi-velocity goes to zero. Several works confirmed the existence of such flat bands at certain ``magic angles" where the electron-electron interactions are maximized \cite{Cao2018, Kerelsky2019}. \\

Yet there are still many open questions concerning  this problem \cite{Cyprian2021,Xueheng2021,Pietro2020,Nguyen2021,GonzalezA2017,2021Ledwidthh}, even in the one-particle operator limit.  For example, the wave function of TBG has been found to be  reminiscent of a quantum Hall wave function in a torus and this opens new analogies to the physics of Landau levels \cite{Uri2020,Hejazi2019, Popov2020,2021Yarden}, the Hofstadter butterfly \cite{Koshino2020,Koshino2013,Oka2021,2021Vogl} or the fractional quantum Hall effect \cite{Kim2013}. There is also an interesting connection to topological phases, Moir\'e edge states and Weyl semimetals \cite{Mikito2020,Manato2021,Pantaleone2021,2021Xiyue,2020Wu,2020Yixing}.\\

Also, as the Moir\'e pattern generates a high electron density localization, interest in making quantum dots with TBG  has been steadily increasing \cite{Mirzakhni2020,2021Yunhua}. Other interesting applications have been found \cite{Fidrysiak2018, Kang2020,Kangjun2019,Vogl2020,Yankowitz2019,Hejazi2019,Christophe2020,Xu2018,Pablo2020,2021Colin}, as well as optical/electrical signatures \cite{Saul2021,Phong2019}.  
The mobility/stability of electrons  is influenced by the triangular geometry of the TBG \cite{Noah2018,GUO2018,Bagchi2020,Park2019}. Previous papers have studied nematicity \cite{Fernandes2021,2021Liu, 2021Kawakami}, phonons/plasmons \cite{Cyprian2021,Young2019}, disorder effects \cite{Hector2021,ochoa2021degradation,Wilson2020} and other important related properties \cite{Guinea2012,Zachary2019_2, Tommaso2020, 2021Ago, 2021AgoHiroki,2021VoPhong,2021Yantao,2021Ochoa}. However, a direct analytic connection with the presence of superconducting phases at magic angles has not yet been achieved completely. As expected,  the interacting behavior of electrons in the Hubbard model is important to characterize the electronic correlations and its fermionic behavior \cite{Noah2018,2021Johannes,2021JieKang,2021Peters,2021Franco}.\\

An important mechanism in the  properties of Moir\'e systems is the  superlattice relaxation \cite{Namm2020,CARR2019,Angeli2018}. This is specially important near AA stacking points, where interlayer hopping tends to be reduced. 
Theoretically, when the hopping that couples layers in AA regions is tuned off the system becomes exactly chiral symmetric \cite{Tarnpolsky2019}. This model shows a recurrence at magic angles and reduces the problem to a  more analytically manipulable Hamiltonian operator. For this reason among others, the chiral Hamiltonian reduces the complexity of the continuum model and captures all the important symmetries and physics of TBG \cite{Tarnpolsky2019}. The mathematical properties and structure  of the wavefunction have been rigorously studied in several works \cite{Popov2020,Zou2018,Mcdonald2021,2021Ledwith, 2021chichinadze}. As one can imagine the graphene layer as two triangular sublattices each one with an equal magnetic flux but with opposite sign, therefore, TBG graphene consists of coupled magnetic fluxes with opposite sign between layers \cite{Khalaf2020,Nori2020,Ledwith2021}. This produces a strong skyrmion behavior in which electrons form vortexes, reflected in the presence of strong electron-electron correlation on specific locations across the Moir\'e superlattice \cite{Zachary2019}. \\

 {\color{blue} To further understand the physics behind the problem, in a previous work we considered the squared Hamiltonian (SH) of the chiral model \cite{Naumis2021}. This represents a renormalization of the TBG that removes one of the bipartites triangular sublattices  for each graphene layer \cite{Naumis2007,Barrios_Vargas_2011,Barrios2013,Naumis2021}. In general, the physics of the SH is the same as that of the original Hamiltonian but the renormalizated operator allows to see properties that in the original model were hidden or difficult to identify. For example, it reveals three physical driving mechanisms: kinetic energy, an effective confinement potential and a non-Abelian gauge field leading to magnetic fields. It also allows to write the Hamiltonian as a simple $2 \times 2$ operator and then use Pauli matrices in which topological properties are more evident. But more importantly, it gives a direct physical interpretation of magic angles in terms of the wave-function geometrical frustration, i.e., we showed that such renormalization folds the spectrum around zero energy and thus zero-mode states correspond to antibonding ground states in a triangular lattice \cite{Naumis2021}. As is well known, antibonding states in triangular lattices are frustrated as the wave-function can not achieve a phase difference of $\pi$ between sites. This cost energy and usually push states into highly degenerate spectral regions and thus to a nearby depletion of states seen as gaps or pseudogaps \cite{Naumis1994,Naumis2002,Barrios_Vargas_2011,Barrios2013}. In Ref. (\cite{Naumis2021}) we showed that magic angles occur whenever the interlayer frustration is exactly zero. Then at magic angles a highly degenerated state is formed and separated by a gap from the rest of the spectrum. Such effect is achieved by a very precise fine tuning  of the wave function Fourier coefficients akin to the Hall effect.  Notice that although previous works  showed some peculiarities about frustration properties \cite{Hridis2019,Nguyen2021, WOLF2019}, it was not clear why such states were at the middle of the band. The same happens with the analytical form of zero-modes, which were identified as reminiscent of a Hall effect ground state without a clear explanation of why the lowest Landau level was found at the middle of the spectrum and not at its bottom end \cite{Tarnpolsky2019,Ledwith2021}.} 

{\color{blue}Yet, several spectral analysis hinted that the first magic angle is differently from others \cite{Tarnpolsky2019,Ledwith2021,WANGG2021}.
For example, numerically it was found that the spectrum of the TBG chiral model shows a remarkable $3/2$ recurrence rule for the magic angles \cite{Tarnpolsky2019}, however, the first angle  does not follows it and the reason is not known. 
As we will see here, their wave functions charge density and phases are remarkably different from others. Thus, it would be very useful to understand the reason of why such behaviors differ from other magic angles. For this reason,  here we present such an  study. Also, this work allows to discern how the physical mechanisms scale between each other as the twist angle is changed.}

The layout of this work is the following. In section \ref{model} we present the model to be studied and the identification of the main physical contributions to the problem. 
Then in section \ref{localization} we study the zero modes wave functions and its localization. In section \ref{Expectation Values}, we study the expectation values of each energy contribution  and discuss the interlayer current contribution, in section \ref{Sec:kpoints} we show why the first magic angle is different from others. Finally, the conclusions are given in the last section.

\section{Squared Twisted Bilayer Graphene chiral Hamiltonian }\label{model}

\begin{figure}[h!]
\includegraphics[scale=0.21]{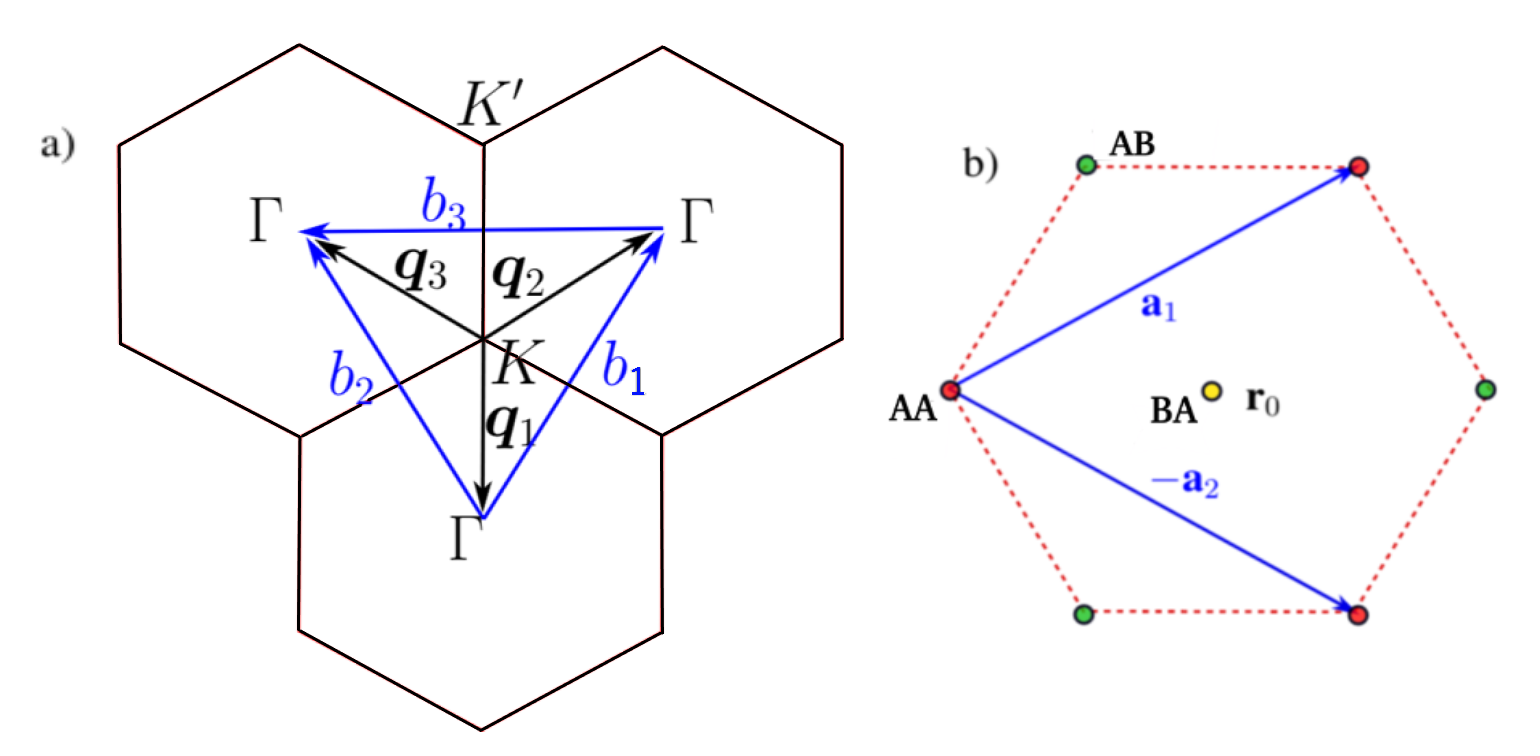}
\caption{ a) Moir\'e Brillouin zones (mBZ) in reciprocal space, $\bm{b_{1,2}}$ are the base vectors. b) Real space Moir\'e unit cell, $\bm{a_{1,2}}$ are two Moir\'e lattice vectors. Point $\bm{r}_0=(\bm{a}_1-\bm{a}_2)/3$ is the BA stacking point where all components of the wave function vanishes at magic $\alpha$.}
\label{fig:moire_uc}
\end{figure}

{\color{blue} The chiral Hamiltonian of twisted bilayer graphene is a variant of the original Bistritzer-MacDonald Hamiltonian in which the $AA$ tunneling is set to zero \cite{Ledwith2021}. We use as basis the wave vectors  $\Phi(r)=\begin{pmatrix} 
\psi_1(r) ,
\psi_2(r),
\chi_1(r),
\chi_2(r)
\end{pmatrix}^T$ where the index $1,2$ represents each graphene layer and $\psi_j(r)$ and $\chi_j(r)$ are the Wannier orbitals on each inequivalent site of the graphene's unit cell. }
The chiral Hamiltonian is given \cite{Tarnpolsky2019,Khalaff2019, Ledwidth2020}, 
\begin{equation}
\begin{split}
\mathcal{H}
&=\begin{pmatrix} 
0 & D^{\ast}(-r)\\
 D(r) & 0
  \end{pmatrix}  \\
\end{split} 
\label{H_initial}
\end{equation}
where the zero-mode operator is defined as, 
\begin{equation}
\begin{split}
D(r)&=\begin{pmatrix} 
-i\Bar{\partial} & \alpha U(r)\\
  \alpha U(-r) & -i\Bar{\partial} 
  \end{pmatrix}  \\
\end{split} 
\end{equation}
and,
\begin{equation}
\begin{split}
D^{*}(-r)&=\begin{pmatrix} 
-i\partial & \alpha U^{*}(-r)\\
  \alpha U^{*}(r) & -i\partial 
  \end{pmatrix}  \\
\end{split} 
\end{equation}

with $\Bar{\partial}=\partial_x+i\partial_y$, $\partial=\partial_x-i\partial_y$. The potential is,
{\footnotesize
\begin{equation}
    U(\bm{r})=e^{-i\bm{q}_1\cdot \bm{r}}+e^{i\phi}e^{-i\bm{q}_2\cdot \bm{r}}+e^{-i\phi}e^{-i\bm{q}_3\cdot \boldsymbol{r}}
\end{equation}
}
where the phase factor $\phi=2\pi/3$ and the Moir\'e lattice vectors are given by $\bm{q}_{1}=k_{\theta}(0,-1)$, $\bm{q}_{2}=k_{\theta}(\frac{\sqrt{3}}{2},\frac{1}{2})$, $\bm{q}_{3}=k_{\theta}(-\frac{\sqrt{3}}{2},\frac{1}{2})$, the Moir\'e modulation vector is $k_{\theta}=2k_{D}\sin{\frac{\theta}{2}}$ with
$k_{D}=\frac{4\pi}{3a_{0}}$ is the magnitude of the Dirac wave vector
and $a_{0}$ is the lattice constant of monolayer graphene, see Fig. \ref{fig:moire_uc}. The physics of this model is captured by the parameter $\alpha$, defined as $\alpha=\frac{w_1}{v_0 k_\theta}$ where $w_1$ is the interlayer coupling of stacking AB/BA with value $w_1=110$ meV and $v_0$ is the Fermi velocity with value $v_0=\frac{19.81eV}{2k_D}$.
{\color{blue} Notice that the Hamiltonian Eq. (\ref{H_initial}) was originally written in Ref. \cite{Tarnpolsky2019} using units where $v_0=1$, $k_{\theta}=1$, thus the operators $\partial$ and $\Bar{\partial}$ are dimensionless. This allows to treat the system with a fixed geometry for any twist angle as in this units $\bm{q}_{1}=(0,-1)$, $\bm{q}_{2}=(\frac{\sqrt{3}}{2},\frac{1}{2})$, $\bm{q}_{3}=(-\frac{\sqrt{3}}{2},\frac{1}{2})$. The twist angle thus only enters in the dimensionless parameter $\alpha$.}

\begin{figure}[h!]
\includegraphics[scale=0.5]{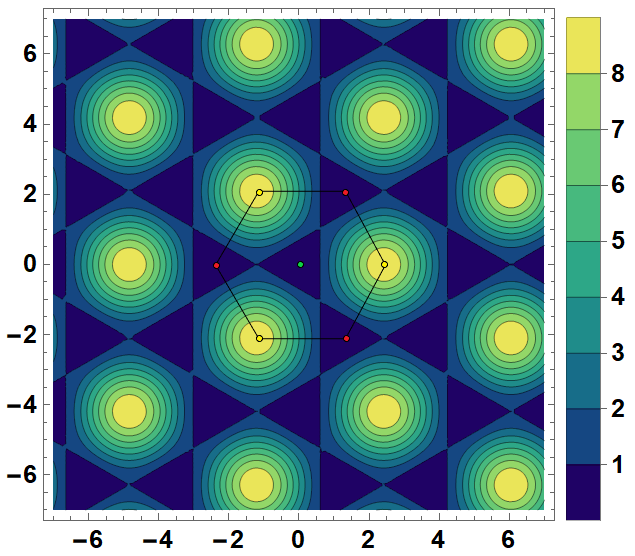}
\caption{Contour plot of the confinement potential $|U(\bm{r})|^{2}$ showing  minima at $AA$ (green) and $AB$ (red) stacking points and maxima at $BA$ stacking points (yellow). As a reference, the Wigner-Seitz cell of the Moir\'e lattice is indicated. For the first magic angle, the $\bm{K}$ wave function tracks such potential.}
\label{fig:Confinement}
\end{figure}

By a renormalization procedure which consists in taking the square of $H$, we found that  \cite{Naumis2021}, 
\begin{equation}\label{eq:H2Def}
\begin{split}
H^{2}&=\begin{pmatrix} -\nabla^{2}+\alpha^{2} |U(\bm{-r})|^{2} & \alpha A^{\dagger}(\bm{r})\\
 \alpha A(\bm{r}) & -\nabla^{2}+\alpha^{2} |U(\bm{r})|^{2}
  \end{pmatrix} 
  \end{split} 
  \end{equation}
The squared norm of the potential is an effective confinement potential,
\begin{equation}
\begin{split}
|U(\bm{r})|^{2} &= 3+2 \cos(\bm{b}_1\cdot \bm{r}-\phi)+2\cos(\bm{b}_2\cdot \bm{r}+\phi )\\
 & +2\cos(\bm{b}_3\cdot \bm{r}+2\phi)   
\end{split}
\end{equation}
where $\bm{b}_{1,2}=\bm{q}_{2,3}-\bm{q}_{1}$ are the Moir\'e Brillouin zone (mBZ)  vectors and $\bm{b}_{3}=\bm{q}_{3}-\bm{q}_2$. In Fig. \ref{fig:Confinement}, we plot $\abs{U(\bm{r})}^{2}$ in real space, with the Wigner-Seitz indicated. This effective potential has an hexagonal   structure where the $r_0$ point in the $BA$ stacking lays in the maximum point of this potential and the minimums lay in the $AA$ and $AB$ stacking points. The off-diagonal terms in $H^{2}$ are,
\begin{equation}\label{eq:Adefinition}
 \begin{split}
A(\bm{r}) & =-i\sum_{\mu=1}^{3}e^{i\bm{q}_{\mu}\cdot\bm{r}}(2\bm{\hat{q}}_{\mu}^{\perp}\cdot \bm{\nabla}-1)\\
  \end{split} 
\end{equation}
and,  
\begin{equation}
 \begin{split}
A^{\dagger}(\bm{r}) & =-i\sum_{\mu=1}^{3}e^{-i\bm{q}_{\mu}\cdot\bm{r}}(2\bm{\hat{q}}_{\mu}^{\perp}\cdot \bm{\nabla}+1)\\
  \end{split} 
\end{equation}
where  $\bm{\nabla}^{\dagger}=-\bm{\nabla}$ with $\bm{\nabla}=(\partial_x,\partial_y)$ and $\mu=1,2,3$. This is an essential point as eigenvalues must be reals (notice that $-A^{\dagger}(\bm{-r})=A(\bm{r})$).
{\color{blue}
 We also define the following operator which contains all the non-diagonal contributions,
\begin{equation}
\begin{split}
\hat{A}(\bm{r}) = 
\begin{pmatrix}
0 & \alpha A^{\dagger}(\bm{r})\\ \alpha A(\bm{r}) & 0
\end{pmatrix}
\end{split} 
 \end{equation}

Also, $\bm{\hat{q}}_{\mu}^{\perp}$ is a set of unitary vectors perpendicular to the set $\bm{q}_{\mu}$,
\begin{equation}
    \bm{\hat{q}}_{1}^{\perp}=(1,0),  \bm{\hat{q}}_{2}^{\perp}=\big(-\frac{1}{2},\frac{\sqrt{3}}{2}\big), 
    \bm{\hat{q}}_{3}^{\perp}=\big(-\frac{1}{2},-\frac{ \sqrt{3}}{2}\big).
\end{equation}

}

The importance of such renormalization is that now we can see the three main ingredients of the problem: i) the kinetic  contribution via the $\grad^{2}$ term (which leads to frustration), ii) a confinement potential $|U(\bm{r})|^{2}$ and, iii) the interlayer interaction  $A(\bm{r})$.

An important feature that we will further analyze is that  $A(\bm{r})$ is made from two terms, and therefore is convenient to define separately the quantities,

\begin{equation}
\begin{split}
 A_{g}(\bm{r})=-2i\sum_{\mu=1}^{3}e^{i\bm{q}_{\mu}\cdot\bm{r}}\bm{\hat{q}}_{\mu}^{\perp}\cdot \bm{\nabla}=2\sum^{3}_{\mu=1}e^{i\bm{q}_{\mu}\cdot\bm{r}} \hat{p}_{\mu} 
\end{split} \label{eq:Agg}
 \end{equation}
and
\begin{equation}
 A_{f}(\bm{r})= -i  \sum_{\mu=1}^{3}e^{i\bm{q}_{\mu}\cdot\bm{r}}
\label{eq:Aff}
 \end{equation}
where the dimensionless projected momentum operators are $\hat{p}_{\mu}=(\bm{\hat{q}^{\perp}}_{\mu}\cdot \hat{\boldsymbol{p}})$, as the dimensionless momentum operator is $\hat{\boldsymbol{p}}=-i\nabla$. {\color{blue} See how Eq.  (\ref{eq:Agg}) is akin to a Lorentz force term.}

Let us also comment some useful symmetries of  $\mathcal{H}$ as they play important roles in the presence of flat-bands \cite{You2019, Zou2018}. For our purposes, the most important symmetry is the exact intravalley inversion symmetry \cite{WANGG2021}, that produce flat-bands and the chirality. The exact intravalley inversion symmetry operator is \cite{WANGG2021},
\begin{equation}
\begin{split}
\mathcal{I}=\sigma_{z}\tau_{y} 
\end{split}
\end{equation} 
{\color{blue} where the $\sigma$ and $\tau$ operators are acting on the sublattice and layer degrees of freedom  respectively and given by two different sets of Pauli matrices \cite{WANGG2021}}. Using such definition we have, 
\begin{equation}
\begin{split}
\mathcal{I}\mathcal{H}(\bm{r})\mathcal{I}^{\dagger}= \mathcal{H}(-\bm{r})
  \end{split} 
  \end{equation} 
Our renormalized Hamiltonian also preserves this symmetry as, 
\begin{equation}
\begin{split}
\mathcal{I}\mathcal{H}^{2}(\bm{r})\mathcal{I}^{\dagger}= \mathcal{H}^{2}(-\bm{r})
  \end{split} .
  \end{equation} 
For the chiral TBG, intravalley inversion follows from the $\mathcal{C}_2\mathcal{T}$ {\color{blue} group symmetry, where $\mathcal{T}$ represents time reversal and $\mathcal{C}_2$ is the cyclic group of order $2$}. The action of $\mathcal{C}_2\mathcal{T}$ is that complex conjugates and exchanges the two sublattices.  In the following section we will further analyze the interplay between the different terms in $H^{2}$ and the role played by the symmetries.

\section{Wave functions: localization properties}\label{localization}

In this section we discuss the localization properties of the zero-modes wave functions for different angles to see if there are differences between the first and higher-order magic angles. We start with the Schrödinger equation $\mathcal{H} \Phi(r)=E \Phi(r)$. Considering only the first spinor component of  $\Phi(r)$, the explicit form of the zero-modes wave function is \cite{Tarnpolsky2019},
\begin{equation} \label{eq:Psik}
\begin{split}
\begin{pmatrix} \psi_{\bm{k},1}(\bm{r}) \\ \psi_{\bm{k},2}(\bm{r})\end{pmatrix}= \sum_{m,n}\begin{pmatrix} 
a_{mn} \\ b_{mn}e^{i\bm{q}_{1}\cdot\bm{r}}
\end{pmatrix} e^{i(\bm{K}_{mn}+\bm{k})\cdot\bm{r}}
\end{split} 
\end{equation}
where $a_{mn}$ and $b_{mn}$ represents the Fourier coefficients  of each spinor component representing layers $1$ and $2$, respectively, and $\bm{K}_{mn}=m\bm{b}_1+n\bm{b}_2$ where $\bm{b}_{1,2}=\bm{q}_{2,3}-\bm{q}_{1}$ are the Moir\'e Brillouin zone  vectors and $\bm{b}_{3}=\bm{q}_{3}-\bm{q}_2$.

For the flat-bands, it has been proved that \cite{Tarnpolsky2019},
\begin{equation}
   \begin{pmatrix} \psi_{\bm{k},1}(\bm{r}) \\ \psi_{\bm{k},2}(\bm{r})\end{pmatrix}=f_{\bm{k}}(z)\begin{pmatrix} \psi_{\bm{K},1} \\ \psi_{\bm{K},2}\end{pmatrix} 
\end{equation}
where $z=x+iy$. $f_{\bm{k}}(z)$ is given in terms of Jacobi theta functions  \cite{Tarnpolsky2019}, or alternatively as a Weierstrass sigma function \cite{WANGG2021}.  
Therefore, the electronic density for layer $j=1,2$ is $\rho_{\bm{k},j}(\bm{r})=|f_{\bm{k}}(z)|^{2}\rho_{\bm{K},j}(\bm{r})$ with $\rho_{\bm{K},j}(\bm{r})=\psi^{\dagger}_{\bm{K},j}(\bm{r}) \psi_{\bm{K},j}(\bm{r})$.

As the analytic form of $f_{\bm{k}}(z)$ is known, our interest here is focused on the study of $\Psi_{K}(\bm{r})$ which corresponds to the ground state of $H^{2}$ at {\it all angles}.

\begin{figure}[h!]
\includegraphics[scale=0.305]{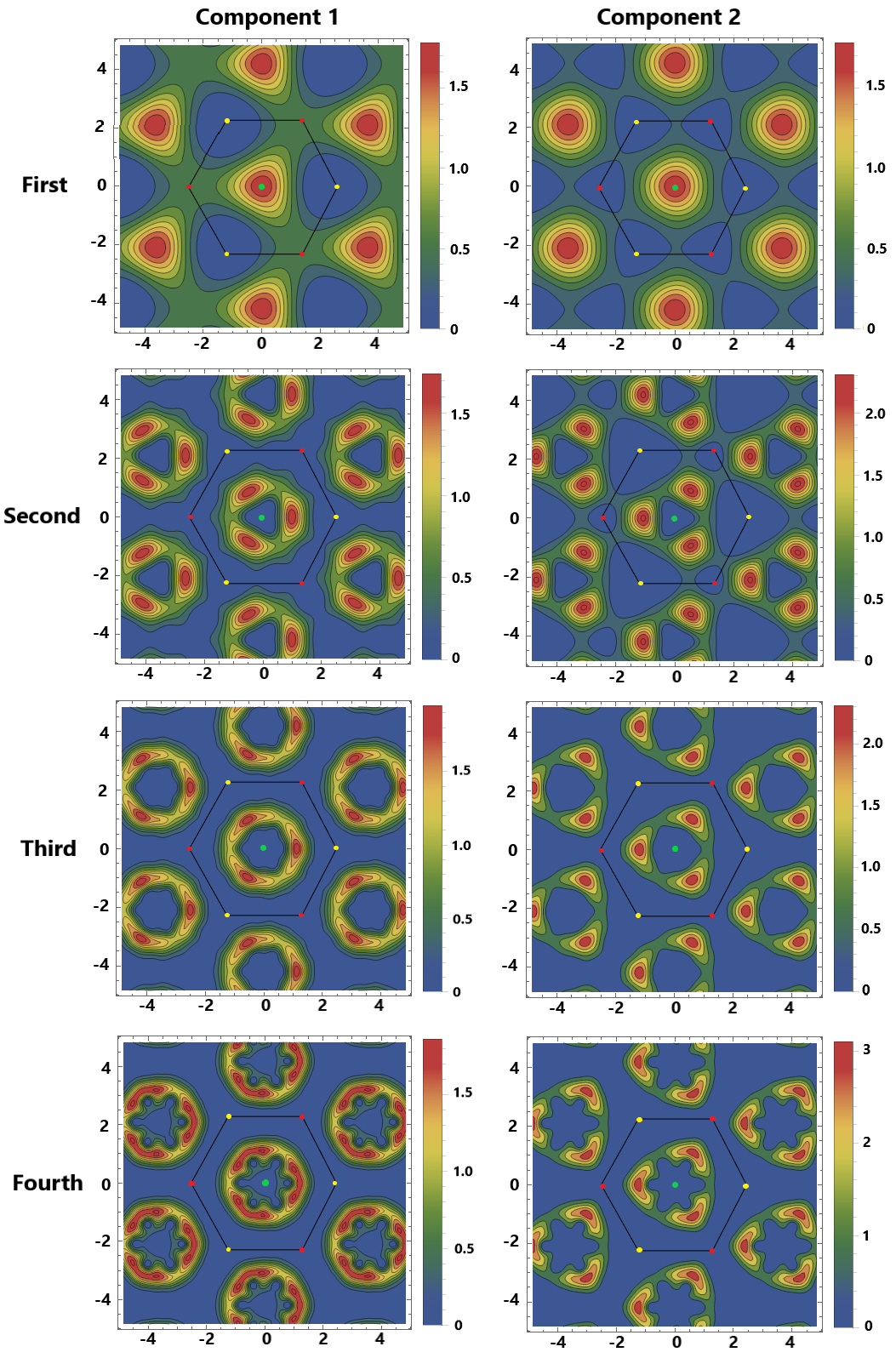}
\caption{Density in real space for the Dirac point $K$ wave function on each layer for the firsts four magic angles $\alpha=0.586, 2.221, 3.751, 5.276$ in the unit cell of the real space representation. The AA (green), AB (red) and BA(yellow) points are indicated. As a reference, the  Wigner–Seitz cell of the Moir\'e lattice is indicated. Notice how the first angle is different from the others as the $AA$ stacking points concentrate the density. }
\label{fig:Density_set1}
\end{figure}

In Fig. \ref{fig:Density_set1} we present the resulting $\rho_{\bm{K}}(\bm{r})$ plots in real space at the first magic angles and for each layer component, obtained by plugging Eq. (\ref{eq:Psik}) into $\mathcal{H}$ to obtain recurrence relations  for the $a_{mn}$'s and $b_{mn}$'s. 
As expected, $\rho_{\bm{K}}(\bm{r})$ present the rotational $C_3$ symmetry. 
Several features are worth noticing, i) the first magic angle is different from the others as the amplitude is centered at the $AA$ stacking points, ii) it tracks the form of the $\abs{U(\bm{r})}^{2}$ confining potential, iii) at other angles the density is confined at somewhat similar locations 
but never at $AA$ points as in the first one.  

At the first magic angle $\alpha_1=0.586$ one can use perturbation theory  \cite{Tarnpolsky2019} to obtain the density (see the Appendix),
\begin{equation}\label{eq:rho1}
\begin{split}
&\rho_{\bm{K},1}(\bm{r})  = 1+\frac{4\alpha^2}{\sqrt{3}}\sum^{3}_{\mu=1}\sin{(\phi+(-1)^{\mu-1}\bm{b}_{\mu}\cdot\bm{r})}\\ 
&+\frac{2\alpha^4}{3} \left.[ 3+2 C(\boldsymbol{r})-\sum^{3}_{\mu=1}\cos{(2\phi+2(-1)^{\mu-1}\bm{b}_{\mu}\cdot\bm{r})} \right. \\
&\left.+2\cos{(2\phi+(-1)^{\mu}\bm{b}_{\mu}\cdot\bm{r})}\right.]
\end{split}
\end{equation}
where $C(\bm{r})=\cos{((\bm{b}_1+\bm{b}_2)\cdot\bm{r})}+\cos{((\bm{b}_1-\bm{b}_3)\cdot\bm{r})}+\cos{((\bm{b}_2+\bm{b}_3)\cdot\bm{r})}$ and the other component,
\begin{equation}\label{eq:rho2}
\rho_{\bm{K},2}(\bm{r})=\alpha^2(3+2\sum^{3}_{\mu=1}\cos{(\bm{b_{\mu}\cdot\bm{r}})}),
\end{equation}

This solution allows to understand the coincidence between Fig. \ref{fig:Confinement} and   the first angle in Fig. \ref{fig:Density_set1} as basically, the $\rho_{\bm{K},2}(\bm{r})$ is just proportional the confinement potential. 

\begin{figure}[h!]
\includegraphics[scale=0.179]{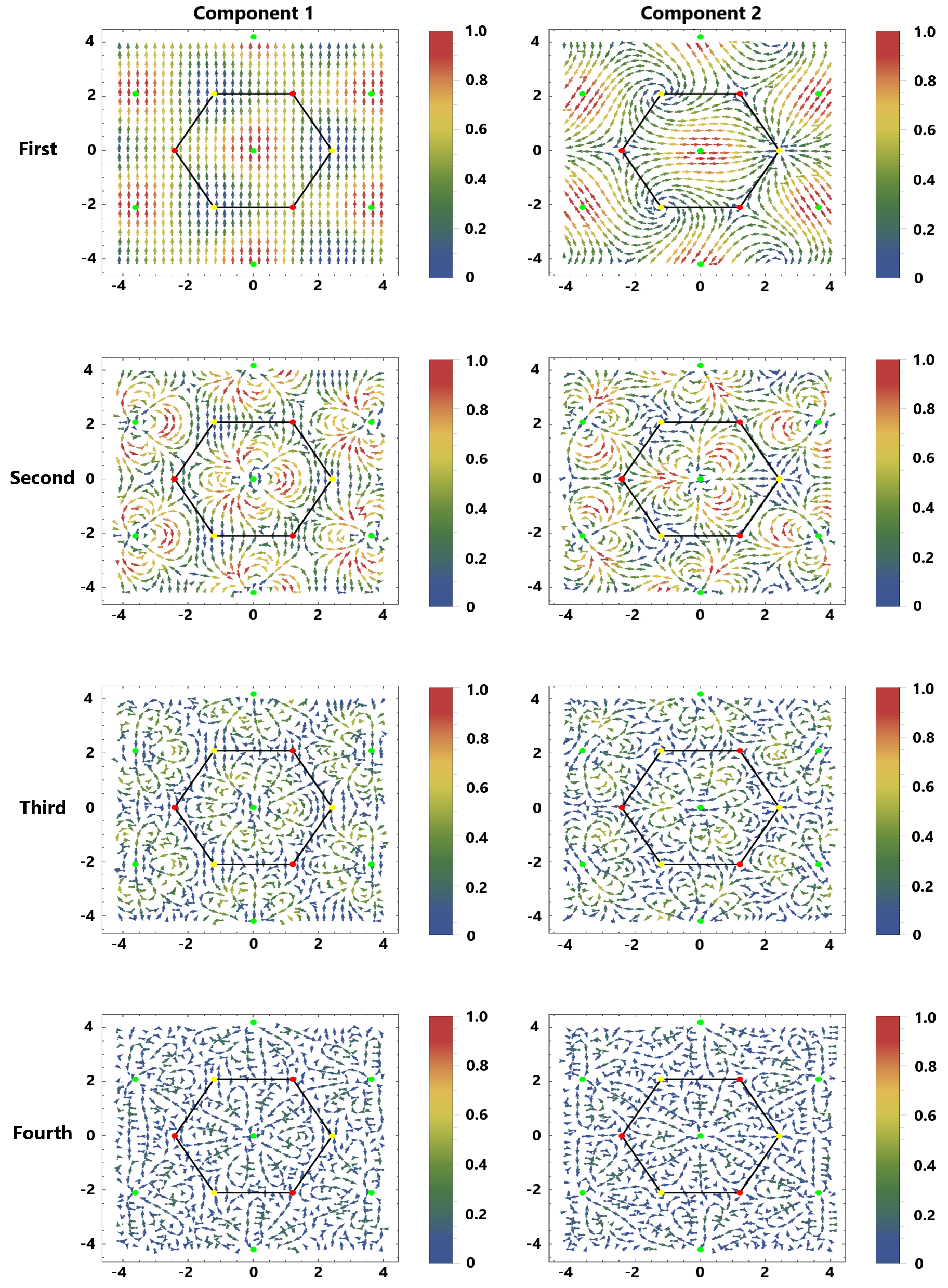}
\caption{Phases in real space where the vertical component of the vectors corresponds to $Im(\psi_{1,2})$ and the horizontal component to $Re(\psi_{1,2})$, in the Dirac point $K$ wave function on each layer, for the firsts four magic angles $\alpha=0.586, 2.221, 3.751, 5.276$ in the unit cell of the real space representation. {\color{blue} The color code is the corresponding wave function amplitude.} The AA (green), AB (red) and BA(yellow) points are indicated. As a reference, the  Wigner–Seitz cell of the Moir\'e lattice is indicated. Notice how the first angle is different from the others as the localization occurs at the $AA$ stacking points.}
\label{fig:Phases}
\end{figure}

In Fig.\ref{fig:Phases}  the real and imaginary parts of the wave function for each layer are shown and the Wigner-Seitz cell is indicated as well. {\color{blue}  The wave functions present  vortices but the most important feature to be seen in Fig.\ref{fig:Phases} is the lack of vortices for the component 1 at the first magic angle, as well as  for the component 2 in the AA stacking point. Such features are in agreement with the perturbative solution for such angle. Although at this moment there are not published figures of the phases to compare with, intralayer currents present vortices \cite{WANGG2021}. However, the vortices of such currents do not coincide with the wavefunction vortices, a feature to be expected since they are made from a sum of different $\bm{k}$ points wavefunctions.  Theoretically it has been suggested that the pairing of the wavefunction vortices are a special signature of the TBG from where superconductivity arises \cite{Khalaf2020}. Here, what is most important for us is the very different behavior of the phases and density associated with the first magic angle when compared with others.}

\begin{figure}[h!]
\includegraphics[scale=0.42]{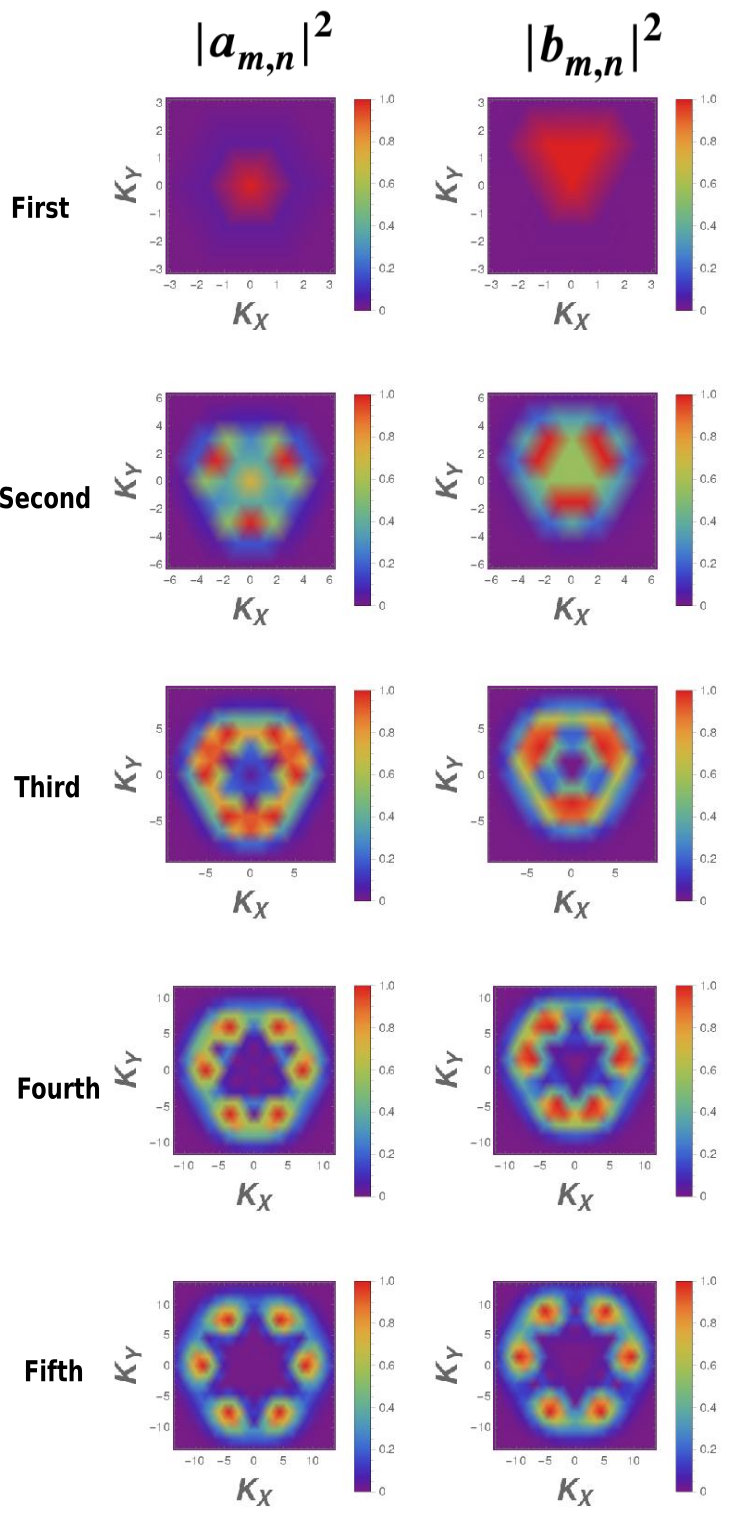}
\caption{Fourier coefficients of the two spinor components for the $\bm{K}$ valley wavefunction. Right and left columns corresponds to the layer $1$ coefficients ($a_{mn}$) and layer $2$ coefficients ($b_{mn}$) respectively. The color represents the amplitude of the coefficient  $|a_{m,s}|^{2}$ or  $|b_{m,s}|^{2}$ in the hexagonal reciprocal lattice centered at location $(k_x,k_y)$. Here we plot their spectral square magnitude for the first four magic angles $\alpha_1=0.586$, $\alpha_2=2.221$, $\alpha_3=3.751$, $\alpha_4=5.276$ and $\alpha_5=6.795$. }
\label{fig:Fourier}
\end{figure}

{\color{blue}To further highlight such differences}, in Fig. \ref{fig:Fourier} we plot the amplitude of the Fourier coefficients $a_{mn}$ and $b_{mn}$ for each layer. Again we see that the first magic angle is remarkably different from  the others, as its main Fourier components contributions are around the origin. However, for the second, third and fourth magic angle there is a hole at $\bm{k} \approx 0$. This hole appears at $\alpha \approx 1$. Also, when $\alpha\rightarrow \infty$, both layers have nearly the same spectral behavior of the Fourier components. 

 The Fourier components of $\Psi(\bm{r})$ in general form complex patterns. However, as seen in Fig. \ref{fig:Fourier}, at the $K,K'$ points most of the coefficients $a_{mn}$, $b_{mn}$ are negligible and the Fourier spectrum consists of six localized peaks forming hexagonal patterns for high values of $\alpha$. Moreover, we find that at the $l$-th magic angle $\alpha_l$, the main contributions of the coefficients $a_{mn}$ are given by  $(m,n)= (\pm l,0),(0,\pm l),(\pm l,\mp l)$. Notably, $l$ has been associated with a Landau-level index \cite{WANGG2021}.

As the wave functions in reciprocal space change but keep a well localized peak, this  means that the localization behavior is far from trivial. To test numerically such observation, here we measure the localization by using a inverse participation ratio \cite{THOULESS1974,Wegner1980,1970BELL} (IPR) ,  
\begin{equation}
 \begin{split}
IPR_{l}(\alpha)=\int_{m}|\psi_{l}(\bm{r})|^{4}d^{2}r
  \end{split} 
\end{equation}
where $l=1,2$ is the index of the top and bottom component of the spinor. 

\begin{figure}[h!]
\includegraphics[scale=0.65]{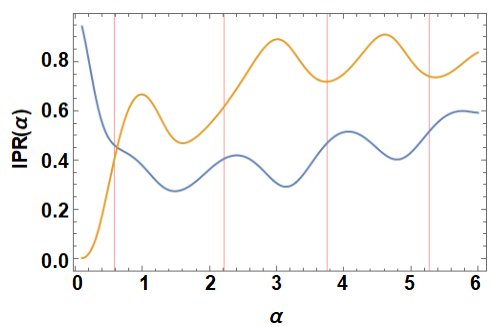}
\caption{IPR as function of $\alpha$ for the top (blue) and bottom (brown) components of the TBG wavefunction at $\bm{k}=\bm{K}$. The red vertical lines indicate the firsts fourth magic angles. Notice how the first-magic angle is different from others as in fact, the IPR of both layers is nearly the same.}
\label{fig:IPR_full}
\end{figure}

In Fig  \ref{fig:IPR_full} we present the IPR for each layer. Analyzing the behavior of the IPR for the top and bottom component. In the limit $\alpha \rightarrow 0$, the IPR reproduces the expected solution $(1,0)$. In the interval $\alpha\in[0,\alpha_1]$ the wavefunction $\psi_1$ becomes less delocalized while $\psi_2$ becomes more localized. When $\alpha$ increases, there is an oscillation in the $IPR_1$ and $IPR_2$. Magic angles occur at inflexion points or at minimas and there is a tendency to increase the overall localization in both layers when $\alpha\rightarrow\infty$.

{\color{blue} Suprisingly, the IPR for magic angles  is the same for all states in the flat band as, except at the poles,
\begin{equation}
|f_{\bm{k}}(z)|^{2}=1.
\end{equation} 
even though $f_{\bm{k}}(z)$ is a complex meromorphic function.
The reason of such relation comes from the normalization of any state corresponding to the flat-band as we must have,
\begin{equation}
   \int_{m} \rho_{\bm{k}}(\bm{r})d^{2}r=\int_{m} |f_{\bm{k}}(z)|^{2}\rho_{\bm{K}}(\bm{r})d^{2}r=1
\end{equation} 
where  $m$ denotes integration over the mBZ.
Then we observe that the wavefunction for $\bm{k}=\bm{K}$ is also normalized as well, from where it follows that in order to be consistent we must have $|f_{\bm{k}}(z)|^{2}=1$ (notice that the poles of $f_{\bm{k}}(z)$ are cancelled out by the zeros of $\psi_{\bm{K}}(\bm{r})$). We have verified numerically that such condition is true.}

To summarize the results of this section, again there are clear signatures in the wave functions that the first magic angle is different from others.

\section{Expectation Values and Interlayer currents}\label{Expectation Values}
To understand the contribution of each physical driving term in the squared Hamiltonian and why the first magic angle is different, we next calculate the expected value of each corresponding  operator in $H^{2}$. From the eigenvalue equation $H^{2}\Psi_{\bm{k}}(\bm{r})=E(\bm{k})^{2}\Psi_{\bm{k}}(\bm{r})$, where $\Psi=(\psi_{\bm{k},1}(\bm{r}),\psi_{\bm{k},2
}(\bm{r}))$, and using Eq. (\ref{eq:H2Def}),

\begin{equation}\label{eq:functional}
\begin{split}
\sum_{j=1}^{2}\big(\langle T_{j} \rangle+\langle V_{j}\rangle \big)+ \langle \hat{A}\rangle=E^{2}(\bm{k})
  \end{split} 
  \end{equation}

where the expected values for a given layer $j=1,2$ are,
\begin{equation}
\langle T_{j}\rangle \equiv -\int_{m}\psi^{\dagger}_{\bm{k},j}(\bm{r})\nabla^{2}\psi_{\bm{k},j}(\bm{r})d^{2}\bm{r} 
\end{equation}
\begin{equation}\label{eq:Vdef}
   \langle V_{j} \rangle \equiv \alpha^{2}\int_{m}|U(\mp\bm{r})|^{2} \rho_{\bm{k},j}(\bm{r})d^{2}\bm{r}
\end{equation}
and  $\langle \hat{A} \rangle=\langle A^{\dagger} \rangle+\langle A\rangle$, with, 
\begin{equation}
\langle A^{\dagger} \rangle \equiv \alpha\int_{m}\psi^{\dagger}_{\bm{k},1}(\bm{r})A^{\dagger}(\bm{r})\psi_{\bm{k},2}(\bm{r})d^{2}\bm{r}
\end{equation}
and 
\begin{equation}
 \langle A \rangle  \equiv \alpha\int_{m}\psi^{\dagger}_{\bm{k},2}(\bm{r})A(\bm{r})\psi_{\bm{k},1}(\bm{r})d^{2}\bm{r}
\end{equation}

{\color{blue} Notice that $\langle A^{\dagger} \rangle$ and $\langle A \rangle$ are not exactly properly defined expected values as involve the bracket of two different wave function components and thus they can be or not complex values. However, the total contribution expected value of the whole off-diagonal terms is real. The total kinetic energy is $\langle T \rangle= \langle T_1 \rangle+\langle T_2 \rangle$ while the total confinement energy is,
\begin{equation}\label{eq:U2bound}
\langle V \rangle=\langle V_1 \rangle+\langle V_2 \rangle \le \frac{8\pi^{2}}{\sqrt{3}}\alpha^{2}.   
\end{equation}
where the last bound is obtained using the wave function normalization.
}

To understand how $\langle \hat{A} \rangle$ depends on the two contributions coming from the off-diagonal terms in $H^{2}$, we will define the space-dependent expected value of the gradient terms in Eq. (\ref{eq:H2Def}) as,
\begin{equation}
\begin{split}
\mathcal{A}_g(\bm{r}) = 
\begin{pmatrix}
\psi^{*}_{1} & \psi^{*}_{2} 
\end{pmatrix}\begin{pmatrix}
0 & \alpha A^{\dagger}_{g}(\bm{r})\\ \alpha A_{g}(\bm{r}) & 0
\end{pmatrix}\begin{pmatrix}
\psi_{1} \\ \psi_{2} 
\end{pmatrix}
\end{split} 
 \end{equation}
which come from the non-diagonal, gradient part, of $H^{2}$,
\begin{equation}
\begin{split}
\hat{A}_g= 
\begin{pmatrix}
0 & \alpha A^{\dagger}_{g}(\bm{r})\\ \alpha A_{g}(\bm{r}) & 0
\end{pmatrix}
\end{split} 
 \end{equation}
Therefore, 
\begin{equation}\label{eq:currents}
\begin{split}
\mathcal{A}_g(\bm{r}) = -2i\alpha\sum_{\mu}\bm{\hat{q}}^{\perp}_{\mu}\cdot(e^{i\bm{q}_{\mu}\cdot\bm{r}}\psi_{1}\nabla\psi^{*}_2+e^{-i\bm{q}_{\mu}\cdot\bm{r}}\psi^{*}_{2}\nabla\psi_1)
\end{split} 
 \end{equation}
By adding its complex conjugate we obtain that,
\begin{equation}
\begin{split}
\mathcal{A}_g(\bm{r})+\mathcal{A}^{*}_g(\bm{r}) &= 2\alpha \frac{m}{e\hbar}\sum_{\mu}\bm{\hat{q}}^{\perp}_{\mu}\cdot(e^{-i\bm{q}_{\mu}\cdot\bm{r}}\bm{j}_{12}
\\ &+e^{i\bm{q}_{\mu}\cdot\bm{r}}\bm{j}_{21})
\end{split} 
\end{equation}
where we defined the interlayer currents as, \begin{equation}\bm{j}_{12}=\frac{ie\hbar}{2m}(\psi_{1}\nabla\psi^{*}_{2}-\psi^{*}_{2}\nabla\psi_{1}) \end{equation}
and
\begin{equation}
\bm{j}_{21}=\frac{ie\hbar}{2m}(\psi_{2}\nabla\psi^{*}_{1}-\psi^{*}_{1}\nabla\psi_{2})
\end{equation}

Such definitions are unusual as involve two different wavefunctions from each layer and only one sublattice. Thus, this requires some comments and thoughts. 
In recent papers, J. Wang et. al \cite{WANGG2021,2021WANG_2,2021WANG_3} proposed a somewhat analogous definition for a second-neighbor intralayer current, i.e., these authors defined,
\begin{equation}
    \bm{j}_{ss}=\frac{ie\hbar}{2m}(\psi_{s}\nabla\psi^{*}_{s}-\psi^{*}_{s}\nabla\psi_{s})
\end{equation}
where $s=1,2$.
Such definition is required as the usual current is obtained from $\partial_x H$ which turns out to be zero in the ground state. Therefore, they defined a current in one of the bipartite lattice as can be seen by performing  a tight-binding calculation of the intralayer orbital current. Here we do not need to appeal to such recourse as already  the square Hamiltonian initially renormalized the hexagonal lattice in a triangular lattice and therefore the second neighborhood interaction is implicit in the renormalization procedure. In Ref. \cite{WANGG2021} it was argued that if $j_{ss}$ is discretized in a tight-binding Hamiltonian, we have that $j_{ss}=i(a^{\dagger}_{s,i}a_{s,j}-a^{\dagger}_{s,j}a_{s,i})$ where $a_j^{\dagger}$ and $a_j$ creates and anhiliates respectively electrons in site $j$.  
In a similar way, if in Eq. (\ref{eq:currents}) we discretize the spinor components gradient we have that $i(\psi_{1}\nabla\psi^{*}_{2}-\psi^{*}_{2}\nabla\psi_{1})\rightarrow i(a^{\dagger}_{2,j}a_{1,i}-a^{\dagger}_{2,i}a_{1,j})$ and the other current component as $i(\psi_{2}\nabla\psi^{*}_{1}-\psi^{*}_{1}\nabla\psi_{2})\rightarrow i(a^{\dagger}_{1,j}a_{2,i}-a^{\dagger}_{1,i}a_{2,j})$. Therefore, this leads to the interpretation of an interlayer current. \\

Let us now integrate over the primitive cell in order to get the expected value of the current. By noting that $\langle \hat{A}_{g} \rangle=\langle \hat{A}_{g} \rangle^{*}$ it follows that, 
\begin{equation}
\begin{split}
\langle \hat{A}_{g} \rangle = 2\alpha\sum_{\mu}\bm{\hat{q}}^{\perp}_{\mu}\cdot \langle \tilde{\bm{j}}_{12}(\bm{q}_{\mu})+\tilde{\bm{j}}_{21}(-\bm{q}_{\mu}) \rangle
\label{intercurrent_operator}
\end{split} 
\end{equation}
where $\tilde{\bm{j}}_{12}(\bm{q}_{\mu})=\int e^{-i\bm{q}_{\mu}\cdot\bm{r}}\bm{j}_{12}(\bm{r})d^{2}\bm{r}$ and $\tilde{\bm{j}}_{21}(-\bm{q}_{\mu})=\int e^{i\bm{q}_{\mu}\cdot\bm{r}}\bm{j}_{21}(\bm{r})d^{2}\bm{r}$. 
Therefore, $\langle \hat{A}_{g} \rangle$ is just the sum of the Fourier components of the current at the points
$\bm{q}_1$,$\bm{q}_2$ and $\bm{q}_3$. In a similar way, the space-dependent expected value of the second ingredient of $\langle A(\bm{r}) \rangle$ is,
\begin{equation}\label{eq:Afhatdef}
\begin{split}
\mathcal{A}_f(\bm{r})= i\alpha\sum_{\mu}(\psi^{*}_{2}\psi_{1}e^{i\bm{q}_{\mu}\cdot\bm{r}}-\psi^{*}_{1}\psi_{2}e^{-i\bm{q}_{\mu}\cdot\bm{r}})
\end{split} 
\end{equation}
where $\mathcal{A}_f(\bm{r})=\mathcal{A}^{*}_f(\bm{r})$, and we define the operator,
\begin{equation}
\begin{split}
\hat{A}_f= 
\begin{pmatrix}
0 & \alpha A^{\dagger}_{f}(\bm{r})\\ \alpha A_{f}(\bm{r}) & 0
\end{pmatrix}
\end{split} 
 \end{equation}

\section{Expectation values and currents at different $k$ points} \label{Sec:kpoints}

\begin{figure}[h!]
\includegraphics[scale=0.85]{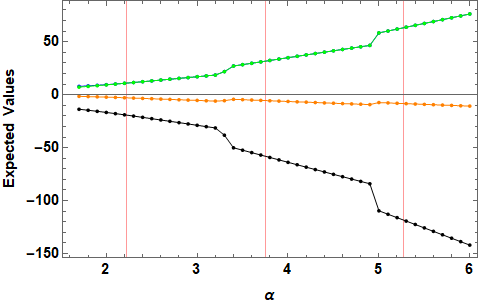}
\caption{{\color{blue}Expected Values in the $\bm{\Gamma}$ point vs $\alpha$. Kinetic Energy $\langle T \rangle$ (Blue), confinement energy $\langle V \rangle$ (Green), $\langle \hat{A}^{\dagger}_{g} \rangle$ (Black) and $\langle \hat{A}^{\dagger}_{f} \rangle$ (Orange). Notice that the kinetic and confinement energies are of the same order, thus is not possible to distinguish the blue curve in this scale. The vertical red lines are the second, third and fourth magic angles. }}
\label{fig:Expected values in GammaBigAlpha}
\end{figure}

{\color{blue} In this section we study all contributions defined in the previous section as a function of the twist at representative points in $\bm{k}$ space. One is the $\bm{\Gamma}$ point which reveals how magic angles arise 
and the other is the $\bm{K}$ point, which is the ground state for all
$\alpha$. Fig. \ref{fig:EgHH} shows such behavior as obtained from the numerical simulation, i.e, the top of the band $E^2(\bm{\Gamma})$ goes to zero at the magic angles while $E^2(\bm{K})$ is the ground state. The $\bm{k}=\zeta$ point, chosen at random in the mBZ, lies inside such interval.

 \subsection{Revealing  the magic angles: $\Gamma$ point expected values}

From Fig. \ref{fig:EgHH} we see that 
$E^2(\bm{\Gamma})$  can be used to reveal the  magic angles as always gives the highest energy of the first $H^{2}$ band. For a flat band to exist, the energy $E^{2}(\bm{\Gamma})$ must be zero.

\begin{figure}[h!]
\includegraphics[scale=0.85]{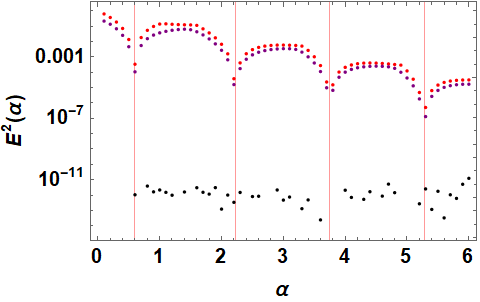}
\caption{$E^2(\bm{k})$ obtained from the squared Hamiltonian Eq. (\ref{eq:H2Def}) at the Dirac $K$ (black), $\Gamma $ (red) and a generic $\zeta $ (purple) point where the red vertical lines indicate the first four magic angles.}
\label{fig:EgHH}
\end{figure}

For the $\bm{\Gamma}$ point it is very illustrative to use  perturbation theory in the limit $\alpha \rightarrow 0$. As shown in the Appendix,
up to linear order in $\alpha$ we have that,

\begin{equation}
     \langle T \rangle=1, \hspace{0.3cm}  \langle V \rangle=0
\end{equation}

\begin{equation}
     \langle \hat{A}_g \rangle=-3\alpha, \hspace{0.3cm} \langle \hat{A}_f  \rangle=-\alpha 
\end{equation}

It follows that,
\begin{equation}
\langle T   +   V \rangle+\langle \hat{A}_f \rangle  +\langle \hat{A}_g \rangle =1-4\alpha \approx E^{2}(\bm{\Gamma})
\end{equation}

In Fig. \ref{fig:ExpectedSmallAlpha} we present a comparison between these expected values and the numerical results showing a good agreement for $\langle T \rangle$, $\langle \hat{A}_f \rangle$, $\langle \hat{A}_g \rangle$   as $\alpha \rightarrow 0$. For $\langle V \rangle$, the agreement is not so good as this requires higher order perturbation terms.
The previous approximation allows to make a crude estimate of the first magic angle as, 
\begin{equation}\label{eq:1stapprox}
    E(\bm{\Gamma}) \approx \pm \sqrt{|1-4\alpha|} 
\end{equation}

Therefore, $\alpha_1 \approx 1/4$, a value below $\alpha_1=0.586$. Higher order terms in the expansion are needed
to increase the accuracy, but yet the main principle behind a magic angle is already present in this simple approach. Further confirmation is provided in Fig. \ref{fig:TVminusAin GammaBiga}
where we show numerically how magic angles arise whenever the curve $\langle T+V \rangle$  intersects $|\langle \hat{A} \rangle|$.

\begin{figure}[h!]
\includegraphics[scale=0.8]{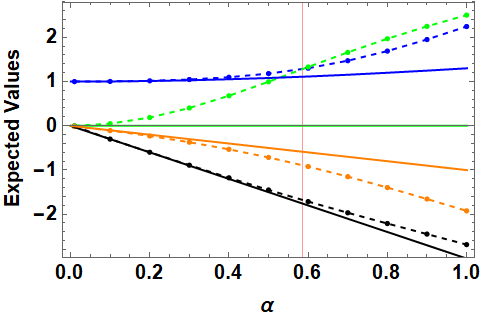}
\caption{Expected values in the $\bm{\Gamma}$ point vs $\alpha$. The numerical results are indicated with dashed lines and points. The kinetic Energy $\langle T \rangle$ is in blue, confinement energy $\langle V \rangle$ (green), $\langle \hat{A}^{\dagger}_{g} \rangle$ (Black) and $\langle \hat{A}^{\dagger}_{f} \rangle$ (Orange). The solid lines are the perturbative solutions (see Appendix). }
\label{fig:ExpectedSmallAlpha}
\end{figure}

Then we conclude  that in going from $\alpha=0$ to $\alpha_1$, the confinement potential starts to contribute and reaches the kinetic energy at the magic angle. The off-diagonal operators always diminish the energy.  As expected, the first magic angle is thus produced when the sum of the kinetic plus confinement energies are equal in magnitude to the expected values of the off-diagonal operators. The particularity here is that for $\alpha_1$ we have $\langle A_g \rangle/\langle A_f \rangle \approx 3$.

\begin{figure}[h!]
\includegraphics[scale=0.95]{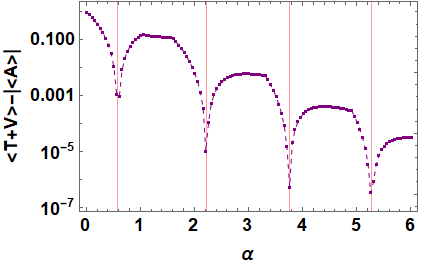}
\caption{{\color{blue} Numerical calculation of $\langle T+V \rangle-|\langle \hat{A} \rangle|$ vs $\alpha$ at the $\bm{\Gamma}$ point for the first four magic angles. For all magic angles $|\langle \hat{A} \rangle|=\langle T+V \rangle$.}}
\label{fig:TVminusAin GammaBiga}
\end{figure}

The numerical results in Fig. \ref{fig:TVminusAin GammaBiga} show  how the same principle applies for other magic angles as  $\langle T+V \rangle-|\langle \hat{A} \rangle|$
goes to zero. However, as seen in Fig. \ref{fig:Expected values in GammaBigAlpha}, for $\alpha>>\alpha_1$ the current term $\langle \hat{A}_{g} \rangle$ dominates over  $\langle \hat{A}_{f} \rangle$, and in fact,  $\langle \hat{A}_{f} \rangle$ can be neglected as we will discuss in the following subsection. Notice also the jumps associated to each $\alpha_n$, possibly related with Landau levels. The other particularity is that $\langle T \rangle \approx \langle V \rangle$ as $\alpha \rightarrow \infty$, thus in  Fig. \ref{fig:TVminusAin GammaBiga} one can not distinguish one from the other in such scale. }

 \subsection{Ground state: $K$ point}

\begin{figure}[h!]
\includegraphics[scale=1.1]{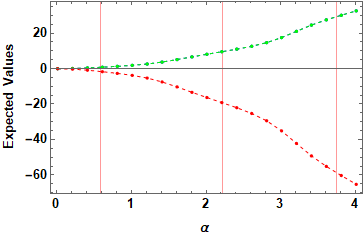}
\caption{Expected value contributions of the energy as function of $\alpha$ at the Dirac point $\bm{k}=\bm{K}$, $\langle T \rangle$ (blue), $\langle V \rangle$ (green) and $\langle \hat{A} \rangle$ (red). As for any angle $\langle T \rangle=\langle V \rangle$, the blue symbols are hidden by the green ones. The conservation of energy implied by Eq. (\ref{eq:budget}) is satisfied as the kinetic, confinement and interlayer contributions always sum zero. }
\label{fig:Expected35}
\end{figure}

The point $\bm{k}=\bm{K}$ is a ground state for any $\alpha$. Let us do some general considerations about it. As $\langle T_{j}\rangle\ge 0$ and $\langle V_{j} \rangle>0$, it follows that for $E^{2}=0$, we have that $\langle A^{\dagger} \rangle \le 0$ and $\langle A \rangle \le 0$. Fig. \ref{fig:Expected35} numerically confirms these results. Other interesting features are seen. The first is already clear from Eq. (\ref{eq:functional}); for the ground state $E^{2}=0$ and due to symmetry we obtain, 
\begin{equation}\label{eq:TVA1}
    \langle T_{1} \rangle+\langle V_{1}\rangle=-\langle A^{\dagger} \rangle
\end{equation}
and,
\begin{equation}\label{eq:TVA12}
    \langle T_{2}\rangle+\langle V_{2}\rangle=- \langle A \rangle
\end{equation}
 The derivation that follows is made by considering a symmetrized basis \cite{Naumis2021}. In this case, there is no way to distinguish the up and low layers except for a relative phase, it follows that we must have
$\langle T_{1}\rangle=\langle T_{2}\rangle$
and $\langle V_{1}\rangle=\langle V_{2}\rangle$. As the general flat-band solutions are given by \cite{WANGG2021},
\begin{equation}
\begin{split}
\Psi(\bm{r})=\begin{pmatrix} 
\psi_1(\bm{r})\\\psi_2(\bm{r})
  \end{pmatrix}=\begin{pmatrix} 
g(\bm{r}) \\ig(-\bm{r})
  \end{pmatrix}\times\Phi_k(\bm{r})
 \label{s_w}
 \end{split} 
  \end{equation}
where $g(\bm{r})$ is a Bloch wave function and $\Phi_k(\bm{r})$ is the quantum Hall wavefunction of the lowest Landau level, we replace such wave function 
into the expressions for $\langle A \rangle$ and $\langle A^{\dagger} \rangle$ to show that, 
\begin{equation}
\begin{split}
\langle A^{\dagger} \rangle= \langle A \rangle.
 \end{split} 
 \label{eq:AAA}
 \end{equation}
This is a reminiscent condition of the intravalley symmetry of  Eq. $\eqref{H_initial}$. 
Therefore, for the total kinetic energy and total confinement energy  we have that,
\begin{equation}
    \langle \hat{A} \rangle=2 \langle A^{\dagger} \rangle=2\langle A \rangle=-\langle T\rangle-\langle V\rangle
    \label{eq:budget}
\end{equation}

Using the bound for the confinement energy, we find that $\langle \hat{A} \rangle$ is bounded by,
\begin{equation}\label{eq:boundforV}
    |\langle \hat{A} \rangle| \le \langle T\rangle+\frac{8\pi^{2}}{\sqrt{3}}\alpha^{2}
\end{equation}

%


\begin{figure}[h!]
\includegraphics[scale=1.1]{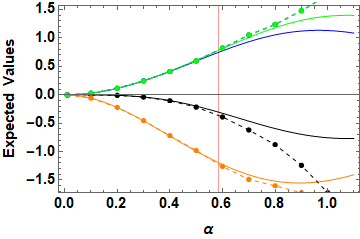}
\caption{{\color{blue}Zoom of the operators expected values versus $\alpha$ for the region $\alpha<<1$.  The filled circles were obtained from the numerical simulation at the Dirac point $\bm{K}=0$, corresponding to: kinetic energy $\langle T \rangle$ (Blue), confinement $\langle V \rangle$ (Green), interlayer current  $\langle \hat{A}_g \rangle$ (Black) and averaged interlayer interaction$\langle \hat{A}_f \rangle$ (Orange). Notice that the numerical data for $\langle T \rangle$ is the same as $\langle V \rangle$ and thus blue circles are not seen. The solid curves were obtained from the analytic perturbative expresions for the operators expected values  up to second order in  $\alpha$ (see Appendix). The same color code as in the numerical data was used for the curves. The red vertical line indicates the first magic angle $\alpha_1=0.586$.}}
\label{fig:ZoomAversusalpha}
\end{figure}

{\color{blue} Fig. \ref{fig:Expected35} further confirms Eq.  (\ref{eq:budget}) and Eq. (\ref{eq:boundforV}). Also, in Fig. \ref{fig:ZoomAversusalpha} we compare the numerical results with the perturbative approach up to second order in $\alpha$ as detailed in the Appendix. The agreement is excellent and allows to: i) further confirm analytically Eq.  (\ref{eq:budget}) in the limit $\alpha \rightarrow 0$ and, ii) test the validity of the numerical approach. }

From the previous results is clear that 
$\langle A \rangle$ will always diminish with $\alpha$ to compensate the increased value of the confinment and kinetic terms. However, such interlayer interaction depends on two terms as in the $\Gamma$ point. This requires a further analysis.


\begin{figure}[h!]
\includegraphics[scale=0.46]{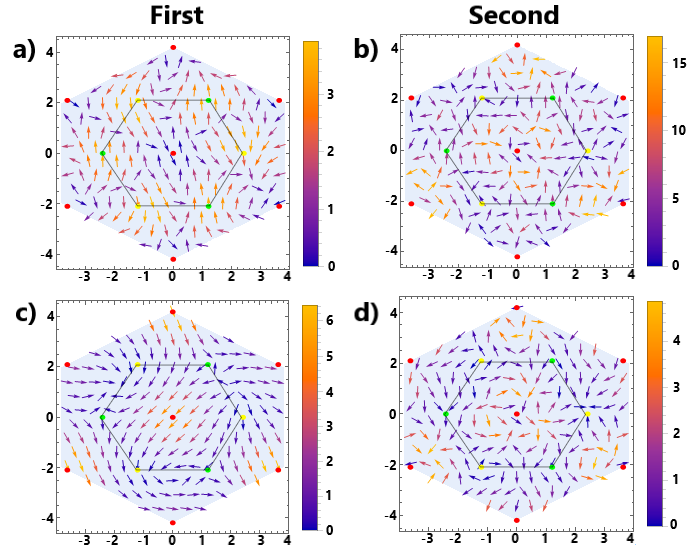}
\caption{Interlayer  contributions in the real space at the Dirac point $\bm{k}=\bm{K}$ for, (a)-(b) corresponds to $\mathcal{A}_g(\bm{r})$ and (c)-(d) corresponds to $\mathcal{A}_f(\bm{r})$ for the two first magic angles $\alpha= 0.586,2.221$. The arrows indicate the direction of the field and the color code the intensity. The stacking points AA (red), AB (green) and BA (yellow) are indicated in the Moir\'e Wigner-Seitz unit cell.}
\label{Interlayer current real space}
\end{figure}

In Fig. \ref{Interlayer current real space} we present  the interlayer current for $\bm{k}=\bm{K}$ in real space. The vectors directions represent polar angle defined by the real and imaginary parts of $\mathcal{A}_f(\bm{r})$ or $\mathcal{A}_g(\bm{r})$. For the first magic angle $\alpha= 0.586$,  $\mathcal{A}_f(\bm{r})$ has more density in the AA stacking point while $\mathcal{A}_g(\bm{r})$ has more current around the BA/AB stacking points. On the other hand, for the second magic angle $\alpha=2.221$, $\mathcal{A}_f(\bm{r})$ and $\mathcal{A}_g(\bm{r})$ have three points of high intensity inside the Wigner-Seitz cell. Also,it is interesting to note that in the AA/AB stacking points there is a vortex behavior as those seen in Fig. \ref{fig:Phases}.

Using the spinor symmetry (\ref{s_w}) and integrating over the primitive cell, it follows that, 
\begin{equation}
\begin{split}
\langle \hat{A}_{f} \rangle&=-2\alpha \Im{\sum_{\mu}\int \psi^{*}_{2}\psi_{1}e^{i\bm{q}_{\mu}\cdot\bm{r}}d^{2}\bm{r}}\\
&=-\alpha\sum_{\mu}Im \left[\tilde{\psi}_2^{*}(q_{\mu})\circledast \tilde{\psi}_1(q_{\mu})\right]
\label{ktheta_operator2}
\end{split} 
\end{equation}
where $\circledast$ means a convolution, $\tilde{\psi_s}$ is the Fourier transform of $\psi_s$.

\begin{figure}[h!]
\includegraphics[scale=0.44]{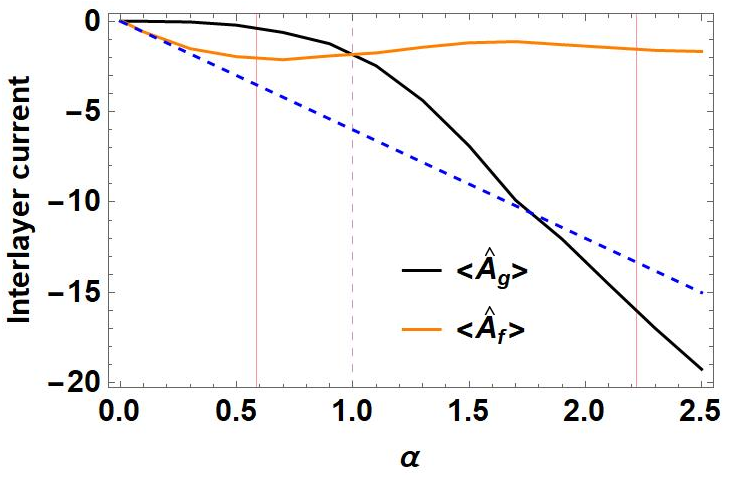}
\caption{Components of the interlayer operator $\langle \hat{A}(\alpha) \rangle$ as a function of $\alpha$ at the  $\bm{K}$ point. The two first magic angles $\alpha= 0.586,2.221$ are indicated with the red vertical lines. At $\alpha \approx 1$ $\langle \hat{A}_g \rangle=\langle \hat{A}_f \rangle$, while the blue dashed line indicates the theoretical lower bound for $\langle \hat{A}_f\rangle \ge -6\alpha$.}
\label{fig:interlayer current}
\end{figure}

In Fig. \ref{fig:interlayer current}, we show the evolution of Eq. (\ref{intercurrent_operator}) and (\ref{ktheta_operator2}). Clearly, for the first magic angle,  $\langle \hat{A}_g \rangle$ and $\langle \hat{A}_f\rangle$ have similar magnitudes but become radically separated after the first magic angle, i.e., $\langle \hat{A}_g \rangle>>\langle \hat{A}_f \rangle$. Therefore, the terms $\langle \hat{A}_f\rangle$ is only relevant for $\alpha < 1 $ making the first magic angle different from others as in the $\Gamma$ point. The reason for such change is easy to see as $\langle \hat{A}_f\rangle$ is bounded by the norm of the wavefunctions and thus, 
\begin{equation}\label{eq:boundAf}
    |\langle \hat{A}_{f} \rangle| \le 6\alpha
\end{equation}
a fact further corroborated by using Eq. (\ref{s_w}) to find its explicit form,
\begin{equation}
    \langle \hat{A}_{f} \rangle=-\alpha\sum_{\mu}\int g^{*}(\bm{r})g(-\bm{r})\abs{\Phi_k(\bm{r})}^{2}\cos{ (\bm{q}_{\mu}\cdot\bm{r}})d^{2}\bm{r} 
\end{equation}

Meanwhile, $\langle \hat{A}_g \rangle$ is proportional to the gradient $\nabla \psi_j$ which is not bounded by $\rho(\bm{r})$. 


\begin{figure}[h!]
\includegraphics[scale=0.31]{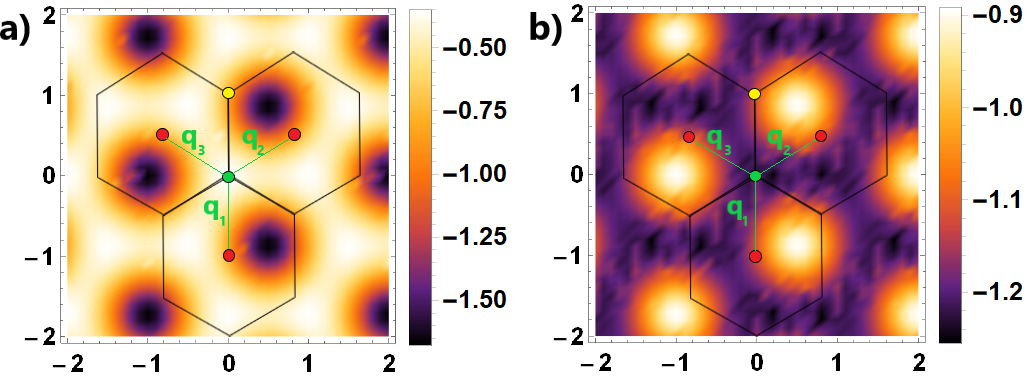}
\caption{Components of the interlayer current mean values as function of momentum. Panel (a) corresponds to  $\langle \hat{A}_g \rangle$ and panel (b) corresponds to $\langle \hat{A}_f \rangle$ at the first magic angle $\alpha= 0.586$. The mBZ is indicated where $\bm{q}_1$, $\bm{q}_2$, $\bm{q}_3$ are the Moir\'e lattice vector and the high symmetry points the $\bm{\Gamma}$ point (red), $\bm{K}^{\prime}$ (yellow) and $\bm{K}$ (green) also are indicated.}
\label{fig:interlayer current momentum space}
\end{figure}

{\color{blue} Further confirmation is obtained by looking at the perturbative solution (see Appendix). In particular $\langle T \rangle=\langle V \rangle \approx 3 \alpha^{2}(1-\alpha^{2})$ while $\langle \hat{A}_g \rangle=-6\alpha^{4}(1+2\alpha^{2})$ and $\langle \hat{A}_f \rangle=-6\alpha^{2}(1+2\alpha^{2})$. Then $\langle \hat{A}_g \rangle/\langle \hat{A}_f \rangle \approx \alpha^{2}$. This ratio goes from zero at $\alpha=0$ to $1$ at $\alpha=1$.} 

\subsection{Comparison between different $\bm{k}$ points}

\begin{figure}[h!]
\includegraphics[scale=0.85]{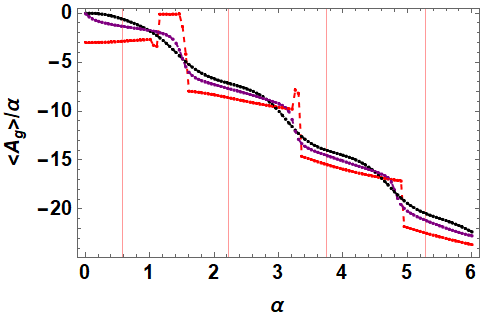}
\caption{{\color{blue}Scaled interlayer current mean values  $\langle \hat{A}_g \rangle/\alpha$ versus $\alpha$ for different representative points in $\bm{k}$ space: $\Gamma$, $K$ and $\zeta$. This last point is chosen at the rim of the black spots of Fig. \ref{fig:interlayer current momentum space}. Notice that $\langle \hat{A}_g \rangle$ is made from two elements: an overall decreasing behavior and at the same time, basins separated by local maxima. Each magic angles is associated with a basin.}}
\label{fig:Agversusalphadifk}
\end{figure}

The previous analysis was made for $\bm{k}=\bm{K}$. In Fig. \ref{fig:interlayer current momentum space} and Fig. \ref{fig:Agversusalphadifk} we extend the analysis  for other flat-band states at magic angles. Fig. \ref{fig:interlayer current momentum space} presents $\langle \hat{A}_g\rangle$ and $\langle \hat{A}_f\rangle$  in  reciprocal space. For the first magic angle $\alpha= 0.586$, the term $\langle \hat{A}_g\rangle$ is maximal where $\langle \hat{A}_f \rangle$ is minimal, both have similar magnitude range. On the other hand, for higher magic angles, the reciprocal space structure of  $\langle \hat{A}_g\rangle$ and $\langle \hat{A}_f\rangle$ preserve the same behavior, however, $\langle \hat{A}_g\rangle$ has a substantially increased magnitude than $\langle \hat{A}_f\rangle$. If $\alpha\rightarrow\infty$, $\langle \hat{A}_g\rangle\gg\langle \hat{A}_f\rangle$. As a consequence, the analysis made for $\bm{k}=\bm{K}$ can be safely extended for all flat-band states.

Therefore,  $\langle \hat{A}_f\rangle$ is only relevant for $\alpha < 1 $ making the first magic angle different from others. {\color{blue}As said before, $\langle \hat{A}_f\rangle$ is limited by $\rho(\bm{r})$ while $\langle \hat{A}_g \rangle$ is proportional to  the wave function gradient. Moreover, as the IPR baseline increases as seen in Fig. \ref{fig:IPR_full}, gradients grow. However, in principle  the overlap between the gradient in one layer and the other layer wave function can diminish.
As $\langle \hat{A}_g \rangle$ is proportional to $\alpha$, to test the gradient effects, in Fig. \ref{fig:Agversusalphadifk} we plot $\langle \hat{A}_g \rangle/\alpha$. This indicates that currents $\bm{j}_{ss}$ due to gradients are the responsible of the effect.  This behavior is also reflected in Fig. \ref{fig:Fourier}, as in the crossover $\alpha \approx 1$ the Fourier coefficients develop a "hole" around $\bm{k}=0$. Notice in Fig.  \ref{fig:Agversusalphadifk} that the magic angles fall inside "basins". Such effect is specially prominent for non high symmetry points as the $\zeta$ point, a fact that will be discussed in a forth coming publication. 
}

Summarizing, the term $\langle \hat{A}_{g} \rangle$ is an energy associated with interlayer current leakage between bipartite sublattices, i.e., at second neighbors. Meanwhile, $\langle \hat{A}_{f} \rangle$ is a weighted average energy associated with overlaps between layers. {\color{blue} The interlayer current magnitude grows as the rotation angle goes to zero, a fact due to the ever increasing spatial gradients of the electron wave function.}

\section{Conclusions}
In this work we presented a theoretical and numerical analysis of the chiral TBG Hamiltonian using a  renormalized Hamiltonian that removes the particle-hole symmetry allowing to identify the main physical elements of the problem and leading to a simple $2\times 2$ matrix operator. Then we studied the electron localization in the TBG. We found numerically that the first magic angle is different from others as the ground state wave function basically tracks the shape of the confinement potential. We calculated the localization using the inverse participation ratio where magic angles are revealed. Interestingly, we proved that all states in the flat band for magic angles have the same participation ratio. We also evaluated the contributions from the kinetic energy, confinement energy and inter-layer interaction for the $\Gamma$ and $K$ points. These contributions were found using perturbation theory and numerically. A good agreement between both was found. Our analysis shows that the $\Gamma$ point reveals how magic angle arises. 
 
 {\color{blue}In particular we found that the first magic angle in the $\Gamma$  point occurs when : 1) the confinement and kinetic energies are the same, 2) the off-diagonal operator is the sum of kinetic and confinement energy, 3) the intralayer current is bigger that the off-diagonal interaction energy term although not negligible. At other magic angles, the balance is dictated only by the kinetic, confinement and interlayer current. Therefore, interlayer currents are the main responsible for bands to shrink. }

 {\color{blue} In other works, the magic angle effects have been associated with the effects of a space dependent magnetic field \cite{Ledwidth2020,WANGG2021}. Our results are in agreement with this idea as the interlayer current can also be interpreted as a Lorentz force too. However, $H^{2}$ allowed us to identify the source of this field: the current between the graphene's underlying triangular sublattices.
 
 It is temping to try to identify our results with some geometrical feature of the twist angle. However, $\alpha$ contains both the geometry and the scale of the energy interaction. Here we found that both are needed in order to make the confinement reach the kinetic energy and produce a strong interlayer current. For very big angles, confinement is just too weak. This is confirmed by the IPR which at the first magic angle has the same value on each layer.} We can also argue that the remarkable $3/2$ rule in the recurrence of $\alpha$ happens in the limit when  $ \langle \hat{A}_g \rangle \gg \langle \hat{A}_f \rangle$, as for higher magic angles the nodal structure of the lowest Landau level does not depend significantly on the Moir\'e unit cell flux. 

\section{acknowledgment}
We thank UNAM-DGAPA project IN102620 and CONACyT project 1564464. Thanks CONACyT schoolarship for providing critical support during the COVID emergency.

\section{Appendix: Perturbative analysis of expected values} 

{\color{blue}

\subsection{$H^{2}$ at the $\bm{\Gamma}$ point}
Consider  the limit $\alpha \rightarrow 0$ for the $\Gamma$ point. The corresponding wave function was  found in Ref. \cite{Tarnpolsky2019}, 
\begin{equation}\label{eq:psiatgamma}
\begin{split}
\psi_{\bm{\Gamma},1}(\bm{r})&=U(-\bm{r})+\frac{\alpha}{3} U(2\bm{r})+\frac{\alpha^{2}}{18}\left((2-e^{i\phi})U(-\sqrt{7}\bm{R}_{\gamma}\bm{r}) \right.\\ & \left.+(2-e^{-i\phi})U(-\sqrt{7}\bm{R}_{-\gamma}\bm{r})-4U(2\bm{r})\right)+...
\end{split}
\end{equation}
and $\psi_{\Gamma,2}(\bm{r})=i\mu_{\alpha}\psi_{\bm{\Gamma},1}(-\bm{r})$, where $R_{\gamma} \bm{r}$ is a counterclockwise rotation on angle $\gamma$ with $\tan(\gamma)= \sqrt{3}/5$ and $\mu_{\alpha}=\pm 1$, the minus sign is used for odd magic angles. The normalization factor is,
\begin{equation}
    N=\sum_{j=1}^{2}\int_{m} \psi_{\bm{\Gamma},j}^{*}(\bm{r})\psi_{\bm{\Gamma},j}(\bm{r})   d^{2}r
\end{equation}

In this case, it is instructive to analyze first how $H^{2}$
operates on the wave function. Consider for example one of the differential equations resulting form Eq. (\ref{eq:H2Def}),  \begin{equation}
    (-\nabla^{2}+\alpha^{2} |U(\bm{-r})|^{2})\psi_{\bm{\Gamma},1}(\bm{r}) + \alpha A^{\dagger}(\bm{r})\psi_{\bm{\Gamma},2}(\bm{r})=E^{2} \psi_{\bm{\Gamma},1}(\bm{r})
\end{equation}
where the square of energy $E^2$ is also expanded in powers of $\alpha$, i.e., $E^2=E_0^{2}+\alpha E_1^{2}+\alpha^{2} E_2^{2}+...$. Next we use the relationship between wave functions on different layers,
 \begin{equation}
 \begin{split}
    (-\nabla^{2}+\alpha^{2} |U(\bm{-r})|^{2})\psi_{\bm{\Gamma},1}(\bm{r}) & \\+  i\mu_{\alpha} \alpha A^{\dagger}(\bm{r})\psi_{\bm{\Gamma},1}(-\bm{r})&=E^{2} \psi_{\bm{\Gamma},1}(\bm{r})
    \end{split}
\end{equation}

If we collect terms up to order $\alpha$ using Eq. (\ref{eq:psiatgamma}), two equations are obtained. The zero order equation is,
\begin{equation}\label{eq:Gamaordezero}
    -\grad^{2} U(-\bm{r})=E_0^{2} U(-\bm{r})
\end{equation}
and at order $\alpha$,
\begin{equation}\label{eq:Gamaordeone}
     -\frac{1}{3}\grad^{2} U(2\bm{r})+i \mu_{\alpha}A^{\dagger}(\bm{r}) U(\bm{r}) =E_0^{2} \frac{1}{3}U(2\bm{r})+E_1^{2} U(-\bm{r})
\end{equation}
From Eq. (\ref{eq:Gamaordezero}) we recover $E_0$ as,
\begin{equation}
    -\grad^{2} U(-\bm{r})=-\grad^{2} \sum_{l=1}^{3} e^{i\bm{q}_l\bm{r}}e^{i(l-1)\phi}=\sum_{l=1}^{3} |\bm{q}_{l}|^{2} e^{i\bm{q}_l\bm{r}}e^{i(l-1)\phi}
\end{equation}
and using $|\bm{q}_{l}|^{2}=1$ we prove that  Eq. (\ref{eq:Gamaordezero}) is indeed true whenever $E_0=1$, in agreement with Ref. \cite{Tarnpolsky2019}. 

Now consider Eq. (\ref{eq:Gamaordeone}). As we did for the order zero component, is easy to show that,
\begin{equation}\label{eq:LaplacianU2}
     -\grad^{2} U(2\bm{r})=4 U(2\bm{r})
\end{equation}
Next we compute $ A^{\dagger}(\bm{r}) U(\bm{r})$ using the two operators $A^{\dagger}_{g}(\bm{r})$ and $A^{\dagger}_f(\bm{r})$. Using the definition for $A^{\dagger}_f(\bm{r})$ and $U(\bm{r})$ we have,
\begin{equation}
    A^{\dagger}_f(\bm{r}) U(\bm{r})=-i \sum_{l,s} e^{-i(\bm{q}_l+\bm{q}_s)\bm{r}}e^{i(s-1)\phi}
\end{equation}
and the sum term is then divided in terms with $l=s$ and $l\neq s$,
\begin{equation}
    \sum_{l,s\neq l} e^{-i(\bm{q}_l+\bm{q}_s)\bm{r}}e^{i(s-1)\phi}+\sum_{l}e^{-i2\bm{q}_l\bm{r}}e^{i(l-1)\phi}
\end{equation}
Using that $\bm{q}_1+\bm{q}_2+\bm{q}_3=0$ and defining a new index  $n=6-(l+s)$,
\begin{equation}
    \sum_{l,s\neq l} e^{-i(\bm{q}_l+\bm{q}_s)\bm{r}}e^{i(s-1)\phi}=-\sum_{n} e^{i\bm{q}_n\bm{r}}e^{i(n-1)\phi} 
\end{equation}
From the definition of $U(\bm{r})$ we finally obtain,
\begin{equation}
     A^{\dagger}_f(\bm{r}) U(\bm{r})= -i(U(2\bm{r})-U(-\bm{r})).
\end{equation}
Let us know consider the  operator $A^{\dagger}_g(\bm{r})$ action. We have,
\begin{equation}
    A^{\dagger}_g(\bm{r}) U(\bm{r})=-i\sum_{l} e^{-i\bm{q}_l\bm{r}} 2 \bm{q}_l^{\perp}\cdot \grad
    \big( \sum_{s}e^{-i\bm{q}_s\bm{r}} e^{i(s-1)\phi} \big)
    \end{equation}
from where,
\begin{equation}
    A^{\dagger}_g(\bm{r}) U(\bm{r})=-i\sum_{l,s\neq l} e^{-i(\bm{q}_l+\bm{q}_s)\bm{r}}
   e^{i(s-1)\phi}  (2 \bm{q}_l^{\perp}\cdot \bm{q}_s)
\end{equation}
Next we use that $ \bm{q}_l^{\perp}\cdot \bm{q}_s=(-1)^{\zeta_{P(l,s)}}\sqrt{3}/{2}$ where $\zeta_{P(l,s)}$ is the sign of the permutation of the indices $l$ and $s$, $+1$ for even and $-1$ for odd. Each pair permutation is obtained from the usual cyclic order $\{1,2,3\}$, so for example,  $\bm{q}_1^{\perp}\cdot \bm{q}_2=\sqrt{3}/2$ while $\bm{q}_2^{\perp}\cdot \bm{q}_1=-\sqrt{3}/2$. Again we use
$\bm{q}_1+\bm{q}_2+\bm{q}_3=0$,
\begin{equation}
    A^{\dagger}_g(\bm{r}) U(\bm{r})=\sqrt{3}\sum_{l,s>l} (-1)^{\zeta_{P(l,s)}}e^{i \bm{q}_n\bm{r}}\big( e^{i(s-1)\phi}-e^{i(l-1)\phi}\big) 
\end{equation}
where $n=6-(l+s)$. Finally,
\begin{equation}
    A^{\dagger}_g(\bm{r}) U(\bm{r})=3i U(-\bm{r}) 
\end{equation}
Then we collect all the previous results inside Eq. (\ref{eq:Gamaordeone}) using $\mu_{\alpha}=-1$ to confirm that the wave vector is an eigenvector of $H^{2}$,
\begin{equation}
\frac{4}{3}U(2\bm{r})-\big(U(2\bm{r})-U(-\bm{r})-3U(-\bm{r}) \big)=E_1^{2}U(-\bm{r})+\frac{1}{3}U(2\bm{r})  
\end{equation}
i.e., comparing terms, we get $E_1^{2}=4$.

The previous analysis confirms that the solutions of $H$ are also eigenfunctions of $H^{2}$. Now we find the expected values of each operator and for the corresponding differential equation. We start with the kinetic energy in the first layer,
\begin{equation}
    \langle T_1 \rangle= -\frac{1}{N}\int_{m} d^{2}\bm{r} \big(U^{*}(-\bm{r})+\frac{\alpha}{3} U^{*}(2\bm{r})\big)\grad^{2} \big( U(-\bm{r})+\frac{\alpha}{3} U(2\bm{r}) \big)
\end{equation}

Using that Eq. (\ref{eq:Gamaordezero}), (\ref{eq:LaplacianU2}) and that
$U^{*}(2 \bm{r})$ and $U(-\bm{r})$ are ortogonal due to symmetry, the contribution of order $\alpha$ is zero,
from where $\langle T_1 \rangle=1/2$. 
Now taking into account the contribution from the equation that results from the second row of Eq. (\ref{eq:H2Def}), we have $\langle T_1 \rangle =\langle T_2 \rangle $. Then,  up to order $\alpha$,
\begin{equation}
     \langle T \rangle= \langle T_1 \rangle+\langle T_2 \rangle=1
\end{equation}
In a similar way, as $U^{*}(2 \bm{r})$  and  $U(-\bm{r})$ are ortogonal,

\begin{equation}
     \langle V \rangle=\langle V_1 \rangle+\langle V_2 \rangle=0
\end{equation}
Finally, the other operators are,
\begin{equation}
     \langle \hat{A}^{\dagger}_f  \rangle=-\alpha 
\end{equation}
while,
\begin{equation}
    \langle \hat{A}^{\dagger}_g \rangle=-3\alpha 
\end{equation}
It follows that,
\begin{equation}
\langle T \rangle  +   \langle V \rangle+\langle \hat{A}^{\dagger}_f \rangle  +\langle \hat{A}^{\dagger}_g \rangle =1-4\alpha \approx E^{2}
\end{equation}
In Fig.
\ref{fig:ExpectedSmallAlpha}
we compare the previous results with the numerical simulation obtaining a good agreement as $\alpha \rightarrow 0$. 

\subsection{$H^{2}$ at the $K$ point}
Here we consider a perturbative solution of Eq. (\ref{pert}) for $\alpha \rightarrow 0$ in the $K$ point as was found in Ref. \cite{Tarnpolsky2019}, 
\begin{equation}
\label{pert}
    \begin{split}
\Psi_{\bm{K}}(\bm{r})=\begin{pmatrix} 
\psi_{\bm{K},1}(\bm{r})\\\psi_{\bm{K},2}(\bm{r})
  \end{pmatrix}=\begin{pmatrix} 
1+\alpha^2 u_2(\bm{r})+\alpha^4 u_4(\bm{r}) \cdots \\\alpha u_1(\bm{r})+\alpha^3 u_3(\bm{r})+\cdots
  \end{pmatrix}
 \end{split} 
\end{equation}

Considering only terms up to order $\alpha^2$ we get:
\begin{equation}
\begin{split}
u_1(\bm{r})&=-i(e^{i\bm{q_1\cdot \bm{r}}}+e^{i\bm{q_2 \cdot \bm{r}}}+e^{i\bm{q_3 \cdot \bm{r}}}) \\ u_2(\bm{r})&=\frac{-i}{\sqrt{3}}e^{-i\phi}(e^{-i\bm{b_1\cdot \bm{r}}}+e^{i\bm{b_2 \cdot \bm{r}}}+e^{-i\bm{b_3 \cdot \bm{r}}})+c.c.
\end{split}
\end{equation}
These functions must be normalized before calculating the expected values. The normalization factor is,
\begin{equation}
    N=\frac{8\pi^{2}}{3\sqrt{3}}(1+3\alpha^{2}+2\alpha^{4}+\frac{6}{7}\alpha^{6}+\frac{107}{98}\alpha^{8}+...)
\end{equation}
With these expresions for $\psi_1$ and $\psi_2 $ we obtain
\begin{equation}
    \int \psi_{2}^{*}(\bm{r})e^{i\bm{q}_u\cdot\bm{r}}\psi_1(\bm{r})d^{2}\bm{r}=\frac{8i\pi^{2}(\alpha+\alpha^{3})}{3\sqrt{3}N}
,\end{equation}
and
\begin{equation}
    \int \psi_{1}^{*}(\bm{r})e^{-i\bm{q}_u\cdot\bm{r}}\psi_2(\bm{r})d^{2}\bm{r}=-\frac{8i\pi^{2}(\alpha+\alpha^{3})}{3\sqrt{3}N}
\end{equation}
for $\mu=1,2,3$. Therefore, taking into account an extra $\alpha$ and $i$ factor from the operator definition, we have,
\begin{equation}\label{eq:perturbativeAf}
     \langle \hat{A}_{f} \rangle \approx \frac{-6(\alpha^{2}+\alpha^{4})}{1+3\alpha^{2}+2\alpha^{4}+\frac{6}{7}\alpha^{6}+\frac{107}{98}\alpha^{8}}
\end{equation}

This result is in agreement with the predicted bound found in Eq. (\ref{eq:boundAf}).  Consider in the same limit the other operator. We have that,
\begin{equation}
    -2i\int  \psi_{2}^{*}(\bm{r})e^{-i\bm{q}_u\cdot\bm{r}}\hat{\bm{q}}^{\perp}_{\mu}\cdot\grad\psi_1(\bm{r})d^{2}\bm{r}=-\frac{8\pi^{2}\alpha^{3}}{3\sqrt{3}N}
\end{equation}
and the terms containg $e^{i\bm{q}_u\cdot\bm{r}} \psi_{1}^{*}(\bm{r})\hat{\bm{q}}^{\perp}_{\mu}\cdot\grad\psi_2(\bm{r})$ give the same result. This confirms Eq.  (\ref{eq:AAA}), i.e., $\langle A \rangle=\langle A^{\dagger} \rangle$. 
By collecting terms we finally obtain that,
\begin{equation}
 \langle \hat{A}_{g} \rangle \approx \frac{-6\alpha^{4}}{(1+3\alpha^{2}+2\alpha^{4}+\frac{6}{7}\alpha^{6}+\frac{107}{98}\alpha^{8})}
\end{equation}

 Next we find the expected values of the kinetic and confining potential operators to confirm the theoretical and numerical analysis. Using that,
\begin{equation}
\int \psi_{1}^{*}(\bm{r}) \grad^{2}\psi_1(\bm{r})d^{2}\bm{r}=-\frac{16\pi^{2}\alpha^{4}}{\sqrt{3}N}
\end{equation}
\begin{equation}
\int \psi_{2}^{*}(\bm{r}) \grad^{2}\psi_2(\bm{r})d^{2}\bm{r}=-\frac{8\pi^{2}\alpha^{2}}{\sqrt{3}N}
\end{equation}
and,
\begin{equation}
\int \psi_{1}^{*}(\bm{r}) |U(\bm{-r})|^{2}\psi_1(\bm{r})d^{2}\bm{r}=\frac{8\pi^{2}(3+2\alpha^{2})}{3\sqrt{3}N}
\end{equation}

\begin{equation}
\int \psi_{2}^{*}(\bm{r}) |U(\bm{r})|^{2}\psi_2(\bm{r})d^{2}\bm{r}=\frac{16\pi^{2}\alpha^{2}}{\sqrt{3}N}
\end{equation}
It follows that,
\begin{equation}
    \langle V \rangle \approx \frac{3\alpha^{2}+6\alpha^{4}+2\alpha^{6}}{1+3\alpha^{2}+2\alpha^{4}+\frac{6}{7}\alpha^{6}+\frac{107}{98}\alpha^{8}}
\end{equation}
and for the kinetic energy term,
\begin{equation}
    \langle T \rangle \approx \frac{3(\alpha^{2}+2\alpha^{4})}{1+3\alpha^{2}+2\alpha^{4}+\frac{6}{7}\alpha^{6}+\frac{107}{98}\alpha^{8}}
\end{equation}
By expanding the denominators we arrive to the final perturbative expectation values,
\begin{equation}
\langle T \rangle \approx 3\alpha^{2}-3\alpha^{4}
\end{equation}
\begin{equation}\label{eq:Vperturba}
      \langle V \rangle \approx 3\alpha^{2}-3\alpha^{4}
\end{equation}
and,
\begin{equation}\label{eq:Afperturba}
     \langle \hat{A}_{f} \rangle \approx 
     -6\alpha^{2}+12\alpha^{4}
\end{equation}
\begin{equation}\label{eq:Agperturba}
     \langle \hat{A}_{g} \rangle \approx
     -6\alpha^{4}+18\alpha^{6}
\end{equation}

in agreement with $ \langle V \rangle= \langle T \rangle>0.$ Also, $\langle V \rangle+\langle T \rangle+\langle A \rangle \approx 0$ up to order $\alpha^{4}$. In Fig. \ref{fig:ZoomAversusalpha} we plot Eqns. (\ref{eq:Vperturba}), (\ref{eq:Afperturba}), (\ref{eq:Agperturba}) and compare them with the numerical results obtained from finding numerically the eigenstates of the Hamiltonian.

\vspace{1cm}
}

\bibliographystyle{plain}

%

\end{document}